\documentclass[conference]{IEEEtran}
\IEEEoverridecommandlockouts
\usepackage{cite}
\usepackage{amsmath,amssymb,amsfonts}
\usepackage{algorithmic}
\usepackage{graphicx}
\usepackage{textcomp}
\usepackage{xcolor}
\usepackage{multirow}
\usepackage{colortbl}

\def\BibTeX{{\rm B\kern-.05em{\sc i\kern-.025em b}\kern-.08em
    T\kern-.1667em\lower.7ex\hbox{E}\kern-.125emX}}

\usepackage{etoolbox}
\makeatletter
\patchcmd{\@makecaption}
{\scshape}
{}
{}
{}
\makeatother

\begin{document}

\title{Who Said What? An Automated Approach to Analyzing Speech in Preschool Classrooms}

\author{\IEEEauthorblockN{Anchen Sun\IEEEauthorrefmark{1},
Juan J Londono\IEEEauthorrefmark{2}, Batya Elbaum\IEEEauthorrefmark{3}, Luis Estrada\IEEEauthorrefmark{2}, 
Roberto Jose Lazo \IEEEauthorrefmark{2},
\\Laura Vitale\IEEEauthorrefmark{2}, 
Hugo Gonzalez Villasanti\IEEEauthorrefmark{4}, 
Riccardo Fusaroli\IEEEauthorrefmark{5, 6},
Lynn K Perry\IEEEauthorrefmark{2},
Daniel S Messinger\IEEEauthorrefmark{2}
\IEEEauthorblockA{\IEEEauthorrefmark{1}Department of Electrical and Computer Engineering\\University of Miami, Coral Gables,
FL, USA}
\IEEEauthorblockA{\IEEEauthorrefmark{2}Department of Psychology\\University of Miami, Coral Gables,
FL, USA}
\IEEEauthorblockA{\IEEEauthorrefmark{3}Department of Teaching and Learning \\University of Miami, Coral Gables,
FL, USA}
\IEEEauthorblockA{\IEEEauthorrefmark{4}Department of Mechanical Engineering\\University of Michigan, Ann Arbor,
MI, USA}
\IEEEauthorblockA{\IEEEauthorrefmark{5}Department of Linguistics, Cognitive Science and Semiotics\\Aarhus University, Nordre Ringgade 1, 8000 Aarhus C, Denmark\\
Email: asun@miami.edu, jjl282@miami.edu, elbaum@miami.edu, lxe300@miami.edu, rjl69@miami.edu,\\ lcv31@miami.edu, hugov@umich.edu, fusaroli@cc.au.dk, lkperry@miami.edu, dmessinger@miami.edu}
}
\\
Codes and Data: https://github.com/DMlabcodeorg/WhoSaidWhat}

\maketitle

\begin{abstract}
Young children spend substantial portions of their waking hours in noisy preschool classrooms. In these environments, children’s vocal interactions with teachers are critical contributors to their language outcomes, but manually transcribing these interactions is prohibitive. Using audio from child- and teacher-worn recorders, we propose an automated framework that uses open source software both to classify speakers (ALICE) and to transcribe their utterances (Whisper). We compare results from our framework to those from a human expert for 110 minutes of classroom recordings, including 85 minutes from child-word microphones (n=4 children) and 25 minutes from teacher-worn microphones (n=2 teachers). The overall proportion of agreement, that is, the proportion of correctly classified teacher and child utterances, was .76, with an error-corrected kappa of .50 and a weighted F1 of .76. The word error rate for both teacher and child transcriptions was .15, meaning that 15\% of words would need to be deleted, added, or changed to equate the Whisper and expert transcriptions. Moreover, speech features such as the mean length of utterances in words, the proportion of teacher and child utterances that were questions, and the proportion of utterances that were responded to within 2.5 seconds were similar when calculated separately from expert and automated transcriptions. The results suggest substantial progress in analyzing classroom speech that may support children’s language development. Future research using natural language processing is under way to improve speaker classification and to analyze results from the application of the automated framework to a larger dataset containing classroom recordings from 13 children and 3 teachers observed on 17 occasions over one year.
\end{abstract}


\section{Background}
\textbf{Overview.} Research on the environments in which children acquire language tends to focus on the home. However, preschool classrooms are also significant environments for language learning. One particular challenge of conducting research in preschool classrooms is that they are noisy environments, containing multiple adult and child speakers. In this paper, we propose a framework for innovative automated measures of linguistic interaction in this naturalistic context. We harness two recent machine learning packages to analyze classroom speech from individual teacher and child microphones. Open source ALICE (Automatic Linguistic Unit Count Estimator) is used to distinguish teacher and child speech; OpenAI’s Whisper is used to transcribe that speech~\cite{gong2023whisper, rasanen2021alice}. We present our computational framework, reliability with a human expert, and describe preliminary results. 

\textbf{Preschool.} In the U.S., the majority of 3- to 5-year old children attend preschool~\cite{FastFact33:online}, making preschool classrooms second only to the home as a context for children’s language and social development~\cite{barnes2017role, ferguson2020social, houen2022eliciting}. Day-to-day preschool experiences promote children’s emerging literacy~\cite{dickinson2011relation} and social skills~\cite{burchinal2008predicting} both directly through children’s interaction with teachers and indirectly through children’s in-the-moment language use~\cite{fasano2023automated, fasano2021granular, mitsven2022objectively}. 

\textbf{Speech Features.} A significant body of research documents the association between high quality features of teacher speech and children’s outcomes~\cite{hadley2023purposes}. In preschool classrooms, the frequency with which teachers use communication-facilitating strategies such as questions is associated with the complexity of the language that children produce~\cite{girolametto2002responsiveness, justice2018linguistic} as well as their vocabulary growth~\cite{cabell2015teacher, perry2018year}. Important features of teacher and child speech include the length of speaker utterances (mean length of utterance in words), the extent to which teachers and children respond to each other’s utterances, and the use of questions and responses by both teachers and children. Teacher questions are of particular importance to children’s language development~\cite{de2005children, massey2008educators} and are the most frequently studied feature of teacher speech~\cite{houen2022eliciting}.
 
\textbf{Measurement.} Measuring speech features is challenging because expert manual transcription and coding of teacher and child speech in noisy, real-life contexts is both time- and labor-intensive. Available technological and human resources have placed limits on the quantity of data collected and the number of speech features that can be measured within the scope of a single study. New sensing technologies and automated measurement of the variables of interest have the potential to transform investigations of how teacher speech affects children's speech, and vice versa, both at moment-to-moment and month-to-month time scales~\cite{elbaum2024investigating,mitsven2022objectively,perry2018year}. Recently, there have been advances in machine-learning recognition of adult wh questions in classroom settings ~\cite{seven2024capturing} and automatic speech recognition has indexed the number of word tokens produced by children~\cite{dutta2022activity, lileikyte2020assessing}. However, understanding the broader content and dynamics of classroom speech has proved elusive.

\textbf{Advances in Speaker Classification and Speech Recognition.} Analyzing classroom speech involves identifying speakers (e.g., as teacher or child) and transcribing their speech. LENA (Language ENvironment Analysis) is a commonly used system for speaker identification~\cite{gilkerson2017mapping}, but open source ALICE is a promising alternative. Automated speech recognition has made promising strides, first with acoustic modeling tools such as the Microsoft Kaldi toolkit~\cite{povey2011kaldi}, the machine learning-based speech emotion recognition for child tantrum simulation~\cite{sun2021multimodal, sun2022multimodal}, and more recently with the emergence of systems like OpenAI’s Whisper.

\textbf{Whisper.} Whisper relies on automatic speech recognition (ASR) through deep learning algorithms, where a neural network is trained on extensive speech samples and their corresponding textual transcriptions~\cite{yu2016automatic, bian2021fedseal}. Whisper appears to have the potential to accurately transcribe spoken language in naturalistic contexts, representing a leap forward in the capabilities of speech recognition technology~\cite{radford2023robust}.  Notably, Whisper’s zero-shot performance with data sets characterized by real-world background noise suggests its robustness~\cite{gong2023whisper}. Whisper and similar systems employ advanced techniques for enhancing accuracy and robustness. These include attention mechanisms that focus on relevant audio segments for better word recognition and methods like noise reduction to cope with challenging audio environments~\cite{li2022recent}. These systems demonstrate the ability to differentiate homophones based on sentence structure and semantics\cite{bianchini2023using}. 

\section{Current Paper}

Methodological advances such as Whisper promise to accurately and speedily transcribe large noisy corpora, thus catalyzing systematic analyses of classroom speech. Here we harness artificial intelligence (AI) in the form of open source machine learning to distinguish and transcribe teacher and child classroom speech. We present a framework in which to simultaneously quantify multiple features of classroom speech quality associated with important developmental outcomes. These features are teacher mean length of utterance (MLU), teacher lexical diversity, proportion of questions in teachers’ and children’s speech, teachers’ responsivity to children’s speech, children’s responsivity to teachers, children’s MLU, and the lexical alignment of teacher and child speech, that is, whether teachers and children reuse any word from their interlocutors in their responses to them~\cite{duran2019align, fusaroli2023caregiver}.

\section{Proposed Framework}

\begin{figure}
    \centering
    \includegraphics[width=0.49\textwidth]{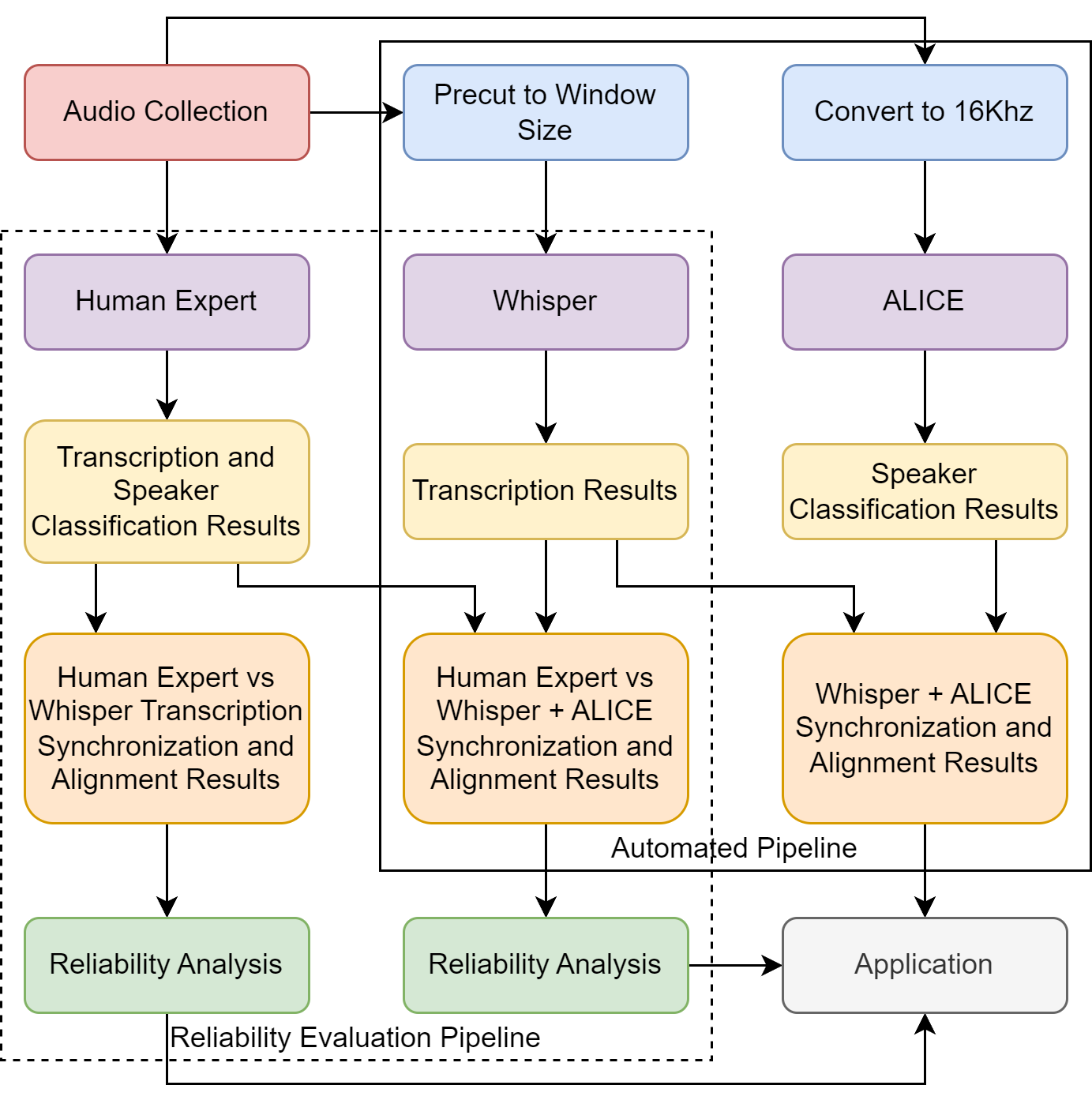}
    \caption{Workflow. Whisper is automated speech-to-text (transcription) software. ALICE provides automated speaker classification (teacher versus child). The human expert provides both speaker classification and transcription. Whisper transcription is used to synchronize ALICE and expert speaker classification.}
    \label{fig:dia}
\end{figure}

In this section, we introduce a comprehensive framework, as shown in Fig.~\ref{fig:dia}, for processing and analyzing large-scale adult-child vocal interactions, leveraging the capabilities of automated language processing tools. We first assess the reliability of the automated pipeline by comparing its output to manual transcription and speaker classification by a human expert. The framework integrates machine learning-based language models for voice transcription (Whisper) and speaker classification (ALICE), which are compared to human expert analysis. In the reliability analysis, we align and synchronize the outputs from the machine learning models with human expert results to evaluate accuracy and consistency. This integrated approach is aimed at providing reliable tools that developmental researchers can use to examine large-scale classroom recordings. To that end, we describe preliminary substantive findings from our reliability analyses.

\subsection{Main Components}

Our workflow integrates three main components. 1) ALICE employs a series of neural networks to distinguish child and teacher vocalizations. 2) Whisper is an opensource audio-to-text neural network developed by OpenAI that was trained on 680K hours of transcribed multilingual speech recorded in multiple environments~\cite{radford2023robust}. It shows impressive performance on datasets used to compare competing algorithmic approaches to speech recognition ~\cite{radford2023robust} and in use with children with developmental disabilities~\cite{o2023automatic}. We use Whisper’s large-v2-model. 3) Results from ALICE and Whisper are rigorously compared to expert speaker classification (child versus teacher) and expert transcription. The overall goal of the framework is processing and aligning large-scale audio recordings as a means of facilitating research on teacher-child communication dynamics in classrooms and other naturalistic settings.

\subsection{Audio Preprocessing}

We collected classroom audio using high-quality recorders worn by children and teachers (see Data Collection). Building on our initial experiences with Whisper, we segmented audio into two-minute epochs before submitting them to Whisper. (Preliminary results using longer epochs yielded a higher error rate in which Whisper duplicated segments during audio that did not contain speech). In addition, if Whisper transcribed a speech segment (a word or series of words) more than two times, only the first and second instances of the speech segment were used in analyses. That is, if Whisper transcribed the speech segment, “Where is it?” three or more times, only two instances of “Where is it?” were used.  This was done to reduce errors in which Whisper transcribed a single speech segment multiple times. ALICE is limited to a frequency input of 16KHz. Therefore, we converted recordings to this frequency before submitting them to ALICE. Whisper analyzed audio at its original 44.1 KHz.

\subsection{Synchronization and Alignment}

In our framework, the results generated by ALICE, Whisper, and the expert include timestamps and must be synchronized and aligned. The framework entails three types of synchronization: Human vs. Whisper, Whisper + ALICE, and Human vs. Whisper + ALICE. Given the importance of Whisper's transcription results and its relatively lower timestamp precision (to the tenth of a second), alignment is based on Whisper's transcriptions for both human expert and ALICE results. If a segment is detected by the human expert or ALICE but not by Whisper, it is added to the final alignment with a blank transcription assignment. Conversely, if Whisper identifies a single segment that corresponds to multiple segments in human expert and ALICE results, these multiple transcriptions are merged, selecting the speaker class that occupies the longest duration. The final aligned results are used for reliability analyses.

\subsection{Overall Reliability Analysis}

For alignment results, we consider the expert coding as the ground truth and compare it with the results from Whisper or Whisper + ALICE.  For human experts, audio clips are manually timestamped and transcribed by an expert coder utilizing the ELAN computer software. For transcription reliability, Whisper and the expert are compared without the need to consider either as ground truth although our experience suggests the expert could be considered ground truth. 

\section{Application}

The proposed framework was implemented using Python 3.9.9 on an Ubuntu 22.04 operating system. The computational resources employed include a machine equipped with a 13900k CPU, 32Gb of RAM, and an RTX4090 24Gb GPU.

\subsection{Data Collection and Processing}

\begin{figure}
    \centering
    \includegraphics[width=0.28\textwidth]{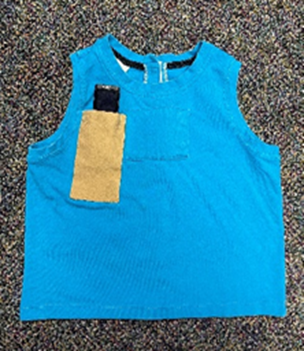}
    \caption{Child vest with Sony recorder (1.43”*4.31”*6.13”) in the front pocket.}
    \label{fig:vest}
\end{figure}

Classroom audio was collected twice per month in a single preschool classroom throughout the school year. Observations (recordings) lasted three hours and twenty minutes on average. Child and teacher audio was captured with individually worn stereo Sony recorders (ICD-UX570 Digital Voice Recorders at LPCM 44.1kHz/16bit quality). Each Sony recorder was worn in a specially outfitted vest worn by child participants or a fanny pack worn by teachers (see Fig.~\ref{fig:vest}). The full dataset contains recordings from 13 children and 4 teachers observed on 17 occasions over the school year. We focus here on an initial dataset compiled quasi-randomly consisting of 110 minutes (25 minutes from teacher recorders, 85 minutes from child recorders) of audio collected from four children (ages 3.8-4.6 years; three girls) and two teachers over two observations (recording days). Informed consent was obtained from the 13 children’s parents and from the teachers. IRB/Ethics approved number:20160509.

\begin{table*}[th]
	\centering
	\begin{tabular}{ccccccc}
		Individual Wearing the Recorder & Duration (min) & Accuracy & Weighted F1 & TPV  & CPV  & Kappa \\
		Child 1                         & 10             & 0.83     & 0.82       & 0.82 & 0.49 & 0.48  \\
		Child 1                         & 45             & 0.74     & 0.74       & 0.74 & 0.74 & 0.47  \\
		Child 2                         & 10             & 0.76     & 0.77       & 0.79 & 0.82 & 0.32  \\
		Child 3                         & 10             & 0.70     & 0.70       & 0.71 & 0.69 & 0.39  \\
		Child 4                         & 10             & 0.76     & 0.76       & 0.77 & 0.87 & 0.46  \\
		Teacher 1                       & 10             & 0.75     & 0.77       & 0.83 & 0.85 & 0.37  \\
		Teacher 1                       & 10             & 0.82     & 0.83       & 0.85 & 0.81 & 0.38  \\
		Teacher 2                       & 5              & 0.80     & 0.81       & 0.84 & 0.75 & 0.20  \\
		Overall                         & 110            & 0.76     & 0.76       & 0.77 & 0.74 & 0.50 
	\end{tabular}
	\caption{Human coding vs Whisper + ALICE. \\ Note. TPV--teacher predictive value--is the proportion of correctly identified teacher utterances (ALICE and expert agreement, true positives, TP) divided by the total number of ALICE-identified teacher utterances (true positives plus false positives [FP]). CPV, child predictive value, is the proportion of correctly identified child utterances (ALICE and expert agreement, true positives) divided by the total number of ALICE-identified child utterances (true positives plus false positives). When identifying teacher utterances, a false negative (FN) occurred when Whisper identified a child utterance when the expert had identified a teacher utterance. The reverse was true for child utterances. F1, calculated separately for teacher and child utterances,  was calculated as 2TP/(2TP+FP+FN). The weighted F1 was the mean of the teacher and child F1 values weighted by the expert-identified number of teacher and child utterances. The “Overall” row is a global calculation in which each utterance is weighted identically. Duration refers to the total amount of time for each clip, including both periods of speech and periods without speech (silence).}
	\label{table:detail}
\end{table*}

\subsection{Speaker Classification Reliability}

\begin{figure}
    \centering
    \includegraphics[width=0.49\textwidth]{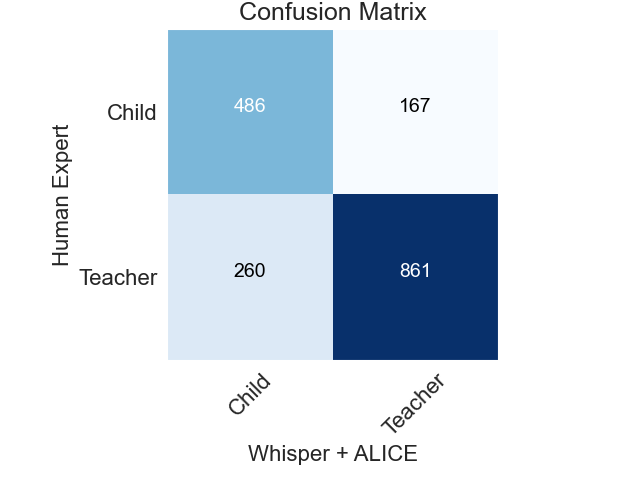}
    \caption{Cross-classification confusion matrix between Whisper + ALICE and the expert. ALICE produced the classification (teacher versus child) while Whisper transcription was used to align results with the expert transcription. Teacher utterances were classified more accurately than child utterances (see TABLE~\ref{table:detail}, Overall, for statistical descriptions).}
    \label{fig:conf}
\end{figure}

TABLE~\ref{table:detail} details metrics assessing the reliability between Whisper + ALICE and the expert in classifying utterances as belonging to children versus teachers. ALICE produced the classification (teacher vs. child) while Whisper transcription was used to align results with the expert transcription. Results are presented per recording and overall for the 110 minutes of recordings. Fig.~\ref{fig:conf} shows the overall cross-classification confusion matrix between Whisper + ALICE and the expert. The overall accuracy (proportion of agreement) was $.76$; Kappa, which corrects for chance agreement, was $.5$. The teacher predictive value (TPV), reflecting agreement on the classification of teacher utterances $(.77)$, was similar to the child predictive value (CPV), which reflects agreement on the classification of child utterances $(.74)$. The weighted F1, $.76$, indexes overall reliability. TABLE~\ref{table:detail} also documents variability in reliability estimates between audio files. The results suggest moderate levels of overall classification.

\begin{figure}
    \centering
    \includegraphics[width=0.49\textwidth]{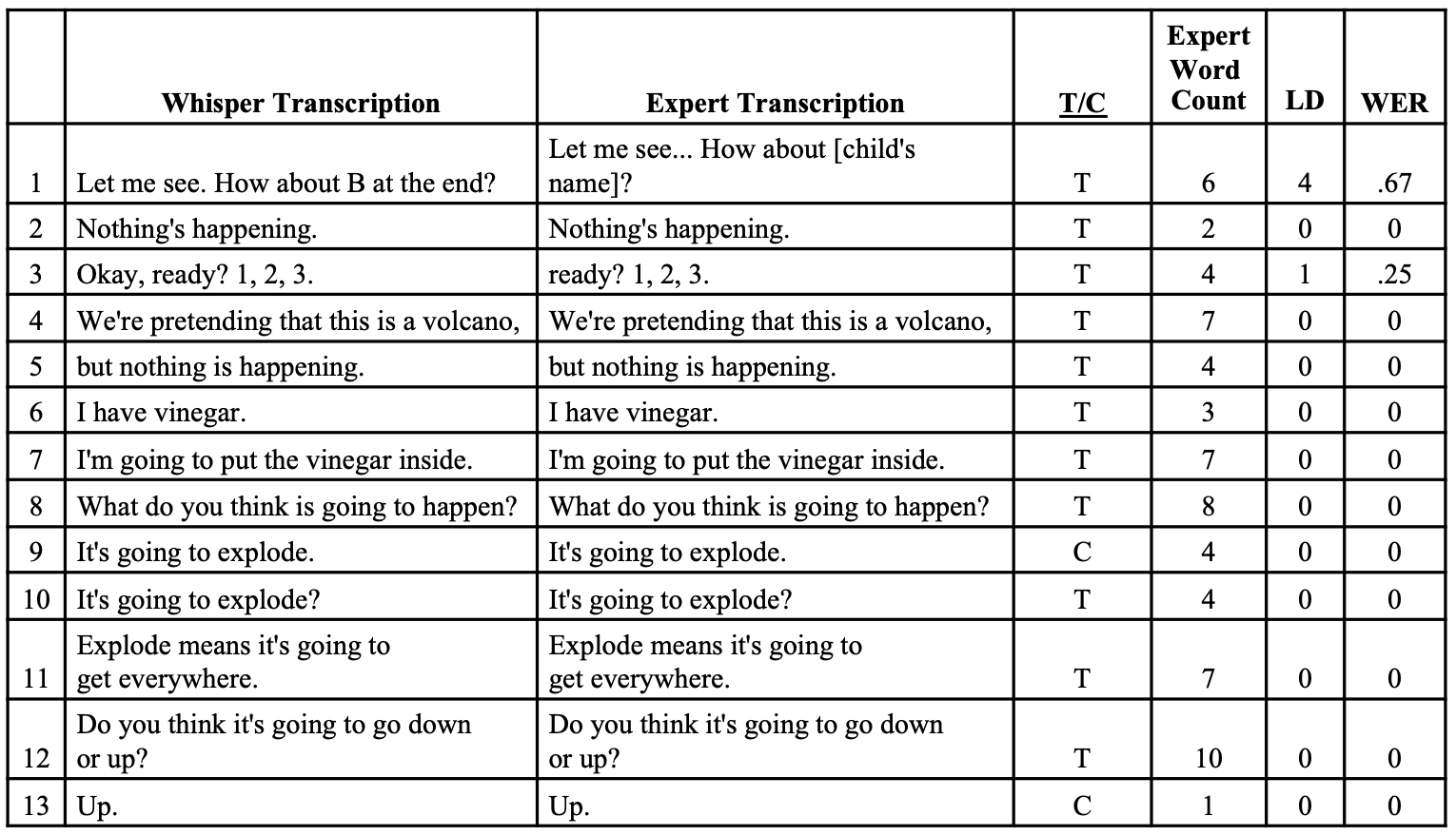}
    \caption{Transcript Comparison. T=Teacher; C=Child; LD=Levenshtein Distance, defined as the number of words that differ between the two transcriptions; WER=word error rate, defined as LD/Expert Word Count. Examples: Line 1: LD=4 based on 1 substituted word and 3 Whisper-added words; WER=4/6=.67. Line 3: LD=1 based on 1 Whisper-added word; WER=1/4=.25.}
    \label{fig:fig4}
\end{figure}

\subsection{Transcription and Speech Feature Reliability}

\begin{table*}[ht]
	\centering
	\setlength\tabcolsep{4pt}
	\begin{tabular}{|clccccccccccc|}
		\hline
		\multicolumn{2}{|c|}{}                                                                                                    & \multicolumn{2}{c|}{\begin{tabular}[c]{@{}c@{}}Child Words \\ (from Child \\ and Teacher \\ Recorders)\end{tabular}}    & \multicolumn{3}{c|}{Child Utterances}                                                                                                                                                                        & \multicolumn{2}{c|}{Teacher Responses}                                                                                                                          & \multicolumn{2}{c|}{\begin{tabular}[c]{@{}c@{}}Proportion of \\ Child Utterances \\ Followed by a \\ Teacher Response\end{tabular}} & \multicolumn{1}{c|}{\begin{tabular}[c]{@{}c@{}}Latency of \\ Teacher to \\ Respond \\ to Child \\ Questions \\ and Non-\\ Questions\end{tabular}} & Alignment                                                                                                                        \\ \hline
		\multicolumn{1}{|c|}{}        & \multicolumn{1}{c|}{Speaker}                                                              & \multicolumn{1}{c|}{MLU}  & \multicolumn{1}{c|}{\begin{tabular}[c]{@{}c@{}}Mean \\ Words \\ per \\ Minute\end{tabular}} & \multicolumn{1}{c|}{\begin{tabular}[c]{@{}c@{}}Total \\ Child \\ Utterances\end{tabular}}   & \multicolumn{1}{c|}{Questions} & \multicolumn{1}{c|}{\begin{tabular}[c]{@{}c@{}}Non-\\ Questions\end{tabular}} & \multicolumn{1}{c|}{\begin{tabular}[c]{@{}c@{}}To \\ Questions\end{tabular}} & \multicolumn{1}{c|}{\begin{tabular}[c]{@{}c@{}}To Non-\\ Questions\end{tabular}} & \multicolumn{1}{c|}{Questions}            & \multicolumn{1}{c|}{\begin{tabular}[c]{@{}c@{}}Non-\\ Questions\end{tabular}}           & \multicolumn{1}{c|}{Mean (sec)}                                                                                                                   & \begin{tabular}[c]{@{}c@{}}Proportion \\ of Question\\ Responses \\ that \\ Yielded No \\ (0) Lexical \\ Alignment\end{tabular}  \\ \hline
		\multicolumn{1}{|c|}{Whisper} & \multicolumn{1}{l|}{\multirow{2}{*}{\begin{tabular}[c]{@{}l@{}}KCHI\\ /CHI\end{tabular}}} & \multicolumn{1}{c|}{3.18} & \multicolumn{1}{c|}{20.51}                                                                  & \multicolumn{1}{c|}{584}                                                                    & \multicolumn{1}{c|}{69}        & \multicolumn{1}{c|}{515}                                                      & \multicolumn{1}{c|}{24}                                                      & \multicolumn{1}{c|}{163}                                                         & \multicolumn{1}{c|}{0.35}                 & \multicolumn{1}{c|}{0.32}                                                               & \multicolumn{1}{c|}{0.25}                                                                                                                         & 0.67                                                                                                                             \\ \cline{1-1} \cline{3-13} 
		\multicolumn{1}{|c|}{Expert}  & \multicolumn{1}{l|}{}                                                                     & \multicolumn{1}{c|}{3.27} & \multicolumn{1}{c|}{21.05}                                                                  & \multicolumn{1}{c|}{702}                                                                    & \multicolumn{1}{c|}{73}        & \multicolumn{1}{c|}{629}                                                      & \multicolumn{1}{c|}{23}                                                      & \multicolumn{1}{c|}{209}                                                         & \multicolumn{1}{c|}{0.32}                 & \multicolumn{1}{c|}{0.33}                                                               & \multicolumn{1}{c|}{0.37}                                                                                                                         & 0.70                                                                                                                             \\ \hline
		&                                                                                           &                           &                                                                                             &                                                                                             &                                &                                                                               &                                                                              &                                                                                  &                                           &                                                                                         &                                                                                                                                                   &                                                                                                                                  \\ \hline
		\multicolumn{2}{|c|}{}                                                                                                    & \multicolumn{2}{c|}{\begin{tabular}[c]{@{}c@{}}Teacher Words\\ (from Child \\ and Teacher \\ Recorders)\end{tabular}}   & \multicolumn{3}{c|}{Teacher Utterances}                                                                                                                                                                      & \multicolumn{2}{c|}{Child Responses}                                                                                                                            & \multicolumn{2}{c|}{\begin{tabular}[c]{@{}c@{}}Proportion of \\ Teacher Utterances \\ Followed by a \\ Child Response\end{tabular}} & \multicolumn{1}{c|}{\begin{tabular}[c]{@{}c@{}}Latency of\\ Child to \\ Respond \\ to Teacher \\ Questions \\ and Non-\\ Questions\end{tabular}}  & Alignment                                                                                                                        \\ \hline
		\multicolumn{1}{|c|}{}        & \multicolumn{1}{l|}{Speaker}                                                              & \multicolumn{1}{c|}{MLU}  & \multicolumn{1}{c|}{\begin{tabular}[c]{@{}c@{}}Mean \\ Words \\ per \\ Minute\end{tabular}} & \multicolumn{1}{c|}{\begin{tabular}[c]{@{}c@{}}Total \\ Teacher \\ Utterances\end{tabular}} & \multicolumn{1}{c|}{Questions} & \multicolumn{1}{c|}{\begin{tabular}[c]{@{}c@{}}Non-\\ Questions\end{tabular}} & \multicolumn{1}{c|}{\begin{tabular}[c]{@{}c@{}}To \\ Questions\end{tabular}} & \multicolumn{1}{c|}{\begin{tabular}[c]{@{}c@{}}To Non-\\ Questions\end{tabular}} & \multicolumn{1}{c|}{Questions}            & \multicolumn{1}{c|}{\begin{tabular}[c]{@{}c@{}}Non-\\ Questions\end{tabular}}           & \multicolumn{1}{c|}{Mean (sec)}                                                                                                                   & \begin{tabular}[c]{@{}c@{}}Proportion \\ of Question \\ Responses \\ that \\ Yielded No \\ (0) Lexical \\ Alignment\end{tabular} \\ \hline
		\multicolumn{1}{|c|}{Whisper} & \multicolumn{1}{l|}{\multirow{2}{*}{FEM}}                                                 & \multicolumn{1}{c|}{4.22} & \multicolumn{1}{c|}{44.48}                                                                  & \multicolumn{1}{c|}{1055}                                                                   & \multicolumn{1}{c|}{258}       & \multicolumn{1}{c|}{797}                                                      & \multicolumn{1}{c|}{87}                                                      & \multicolumn{1}{c|}{166}                                                         & \multicolumn{1}{c|}{0.34}                 & \multicolumn{1}{c|}{0.21}                                                               & \multicolumn{1}{c|}{0.32}                                                                                                                         & 0.78                                                                                                                             \\ \cline{1-1} \cline{3-13} 
		\multicolumn{1}{|c|}{Expert}  & \multicolumn{1}{l|}{}                                                                     & \multicolumn{1}{c|}{4.46} & \multicolumn{1}{c|}{47.01}                                                                  & \multicolumn{1}{c|}{1159}                                                                   & \multicolumn{1}{c|}{286}       & \multicolumn{1}{c|}{873}                                                      & \multicolumn{1}{c|}{106}                                                     & \multicolumn{1}{c|}{211}                                                         & \multicolumn{1}{c|}{0.37}                 & \multicolumn{1}{c|}{0.24}                                                               & \multicolumn{1}{c|}{0.47}                                                                                                                         & 0.88                                                                                                                             \\ \hline
	\end{tabular}
	\caption{Teacher and Child Speech Features and Responses from Whisper and Expert Transcriptions \\ Note. MLU is the mean number of words per utterance. KCHI: key child wearing the recorder. CHI: a child other than the key child. FEM: female adult (assumed teacher). All teacher and child speech features and responses were generated from both teacher and child recorders. Both Whisper and expert speaker classification was provided by the expert coder.}
	\label{table:2}
\end{table*}

\subsubsection{Word Error Rate} We calculated word error rate (WER) separately for data from teachers’ and children’s recordings. Teacher WER was calculated on audio from teacher microphones in which an adult was the identified speaker. Child WER was calculated on audio from child microphones in which the child wearing the microphone was the identified speaker. Word Error Rate is the most rigorous metric for comparing two transcripts. WER is calculated based on the Levenshtein Distance, which quantifies the difference between two utterances by counting the number of words needed to transform one into the other (see Fig.~\ref{fig:fig4}). If Whisper or the expert transcribed an utterance that was not transcribed (i.e., was entirely omitted) by the other, the WER is the total number of words in the transcribed utterance (i.e., all the words are treated as error). The mean WERs were $.147$ for teacher speech and $.150$ for child speech. This means that $85\%$ of words (for both teachers and children) were identical in the expert and Whisper transcripts. This is a relatively high level of transcription accuracy and is consistent with previous research with children~\cite{o2023automatic}.

\begin{table}[h]
	\setlength\tabcolsep{4pt}
	\begin{tabular}{lccc
			>{\columncolor[HTML]{FFFFFF}}c 
			>{\columncolor[HTML]{FFFFFF}}c }
		& MLU  & Questions & \begin{tabular}[c]{@{}c@{}}Non-\\ Questions\end{tabular} & \begin{tabular}[c]{@{}c@{}}Proportion of \\ Responses \\ to Questions\end{tabular} & \begin{tabular}[c]{@{}c@{}}Proportion of \\ Responses to \\ Non-Questions\end{tabular} \\
		Child   & 0.84 & 0.94      & 0.75                                                     & 0.74                                                                               & 0.97                                                                                      \\
		Teacher & 0.93 & 0.97      & 0.95                                                     & 0.93                                                                               & 0.76                                                                                     
	\end{tabular}
	\caption{Intraclass Correlations \\ Note. Intraclass correlations index the absolute reliability between speech features calculated from automated and expert transcriptions over different audio files based on expert speaker identification. Questions and non-questions were measured as a rate per minute. For the child row, we report the proportion of teacher responses to child questions. For the teacher row we report the proportion of child responses to teacher questions.}
	\label{tab:3}
\end{table}

\subsubsection{Speech Feature Comparison Overview} We next compared features of classroom speech that were quantified (transcribed) by Whisper and an expert, both based on expert identification of who was speaking. That is, we not only assessed the reliability of Whisper’s word-by-word transcription, but we also compared Whisper and expert quantification of specific features of child and teacher language use. All recordings were used in these analyses, that is, a child’s speech features were derived from audio from both teacher and child recorders, and the same was true of teacher speech features. This aided in capturing teacher and child responses to one another’s utterances. We compared Whisper and expert measures of teacher and child utterances (words per minute, mean length of utterance, and number of utterances), questions and responses, and the presence of lexical alignment in those responses. The utterance was the primary unit of analysis in calculating means, standard deviations, and correlations (see TABLE~\ref{table:2}). Intraclass correlations using audio files as the unit of analysis are contained at the end of this section (see TABLE~\ref{tab:3}).

\subsubsection{Mean Length of Utterances in words (MLU)} For these analyses,  if either Whisper or the expert did not identify an utterance, the length of the utterance in words was calculated as $0$. That is, omissions of utterances were penalized. Mean MLU for teachers was $4.22 (SD=3.23)$ for Whisper and  $4.46 (SD=3.13)$ for the expert. The mean MLU for children was $3.18 (SD=2.17)$ for Whisper and  $3.27 (SD=2.60)$ for the expert. The correlation between Whisper and expert MLU was $r=.87$ for teachers and $r=.80$ for children. The results suggest reasonable correspondence between Whisper and the human expert in estimates of speech complexity based on MLU (see TABLE~\ref{table:2}).

\subsubsection{Speech Rate} Using features from both Whisper and expert transcriptions, teacher MLU was about one-third larger than child MLU. However, both Whisper and the expert indicated that teachers also spoke at more than double the rate of children (words per minute; see TABLE~\ref{table:2}), producing more utterances than children. Note that Whisper suggested a somewhat larger ratio of teacher to child utterances $(1.81 [1055/584])$ than the expert $(1.65 [1159/702])$.

\subsubsection{Questions and Non-questions} We next calculated agreement between expert and Whisper on the occurrence of question versus non-question utterances for instances in which both expert and Whisper identified an utterance. We identified questions based on the occurrence of a question mark (both Whisper and the expert used question marks). The percentage of agreement between expert and Whisper on the occurrence of a question was $.96$ for both teachers and children. The corresponding Cohen’s Kappas, which correct for chance agreement, were $.83$ for children and $.88$ for teachers. Both Whisper $(69/584=.12)$ and the expert $(73/702=0.10)$ indicated that approximately one-tenth of child utterances were questions while approximately one-fourth of teacher utterances were questions (Whisper: $258/1055=0.24$ and expert $286/1159=0.25$). Thus there was strong agreement between Whisper and the expert on the occurrence of question versus non-question utterances (and their relative proportions) for children and teachers. 

\subsubsection{Responses} We defined child responses to teacher utterances (both questions and non-questions) as instances of a child utterance that was the first utterance to follow a teacher utterance and was within 2.5 seconds of the preceding utterance, where 2.5 seconds amply covers child utterances that are responses to a prior teacher question based on previous research~\cite{casillas2016turn, nguyen2022systematic}. We identified teacher responses to children’s utterances using the same method. Note that all responses are also utterances. Analyses based on both Whisper and expert transcriptions indicate that approximately one-third of child utterances (both questions and non-questions) were responded to by teachers (see TABLE~\ref{table:2}). Likewise, in both Whisper and expert transcriptions, children responded to teacher questions about one-third of the time but responded to teacher non-questions in approximately one-fifth (Whisper) and one-fourth (expert) of cases. These results indicate a promising correspondence in the rate of utterances that elicited responses--the heart of linguistic interaction--between speech features derived from the automated framework and expert.

\subsubsection{Lexical Alignment} Finally, we calculated the proportion of responses that contained none of the words in the preceding utterance ($0$ alignment). A high proportion of both teacher (Whisper=$.78$, expert=$.88$) and child responses (Whisper=$.67$, expert=$.70$) had no lexical overlap with the preceding utterance. All words in a pair of utterances are used to calculate lexical overlap. Proportions of responses without lexical overlap for child responses appeared larger for the expert than for Whisper. This may be because Whisper relies on linguistic context in its transcriptions, and may be more likely to transcribe a child utterance as containing words that it also detected in the preceding adult utterance than is the expert (increasing lexical overlap and decreasing the proportion of child responses with no lexical overlap). By contrast, there was a close match between expert and Whisper on the absence of lexical alignment in teacher responses to children.

\subsubsection{Intraclass Correlation Approach} For more comprehensive reliability evaluations, we conducted absolute level intraclass correlations using audio files as the unit of analysis (see Table~\ref{tab:3}). Each audio file reflects the language environment of a given teacher or child. In reliability theory parlance, audio files are the objects of measurement and variance between audio files is true variance. These intraclass correlations reflect audio file variance as a proportion of total variance. Total variance contains two sources of error--variance due to differences in measurement (Whisper versus expert), and error variance.

\subsubsection{Intraclass Correlation Results} Overall, intraclass correlations ranged between $.74$ and $.97$. Intraclass correlations for the rate of child question asking, the proportion of child questions that were responded to, and the proportion of teacher non-questions that were responded to by the child ranged from $.74-.76$, indicating only moderate reliability on these features. All other intraclass correlations--including child and teacher MLU and rate of question asking--were high ($.84-.97$). The results indicate that a substantial quantity of the variance in measured speech features was associated with differences in those features between audio clips.

\section{Discussion}

Preschool is an important developmental context for the majority of 3- to 5-year olds in the United States. Within preschools, there is pervasive evidence that teacher speech features including question use and MLU are associated with children’s language outcomes~\cite{hadley2022meta,walsh2018we}. Likewise, children’s own speech production may be a bridge between teacher speech complexity and child language outcomes~\cite{mitsven2022objectively}. Finally, language interactions that occur in preschool settings may have an outsize impact on children whose families live in poverty as well as children with developmental disabilities. 

Manual speaker classification and transcription is often a limiting factor in understanding classroom speech. For example, expert transcription of the $110$ minutes of audio data reported here took approximately $55$ hours ($5$ hours of expert transcription per 10 minutes of audio). Researchers have begun to tackle this problem through automated quantification of selected speech features from classroom audio~\cite{dutta2022activity,lileikyte2020assessing,seven2024capturing}. We add to this literature with an automated framework for the large-scale analysis of speaker classification and transcription (who said what).

Speaker classification results indicated moderate levels of overall reliability for both teacher and child utterances. Word error rates--a rigorous metric--indicated relatively high levels of transcription accuracy for both teacher and child recordings. Comparison of expert and automated processing was not limited to standard reliability metrics such as word error rate, but was extended to the speech features that are the likely substantive areas of analysis for classroom interaction research. These analyses used all available audio from both teacher and child recorders. The results suggest promising levels of correspondence on teacher and child MLU, rate of speech, use of questions, and responses to questions. However, each of these features must be examined individually. For example, our data suggest that Whisper over-estimated lexical alignment between teacher utterances followed by child utterances. This may be due to the tendency of this large language model to “hear” words in child utterances that had been identified in the previous teacher utterances.

Although descriptions of speech features are preliminary, both the automated framework and expert coding suggest that teachers have somewhat higher MLU than children and produce many more utterances than children. Teachers asked more questions than children but both teachers and children responded to about one-third of their interlocutor’s questions. The proportion of responses that contained words that were also contained in the preceding utterance was low. It should be noted, however, that interchanges such as those in Fig.~\ref{fig:fig4} (teacher: “What do you think is going to happen?” child: “it's going to explode”) contain no lexical overlap. More sophisticated analyses of alignment--including semantic alignment based on meaning--are required~\cite{fusaroli2023caregiver}.

The current research has multiple limitations. Speaker classification accuracy requires improvement and current results were based on expert speaker classification. Machine learning of Whisper transcriptions to bolster speaker classification is one promising direction of research. Transcription results were relatively accurate but were limited to the teacher or child wearing the recorder, suggesting that the distance between mouth and recorder microphone might be a crucial variable. However, speech feature results showed impressive correspondence between automated and expert measurement both at an utterance level and an audio recording level. 

\section{Conclusion}
The current results suggest a framework for advancing the automated analysis of classroom speech. High-quality recorders worn by teachers and children yielded reliable automated transcription of what individuals said. Moreover, there were high levels of agreement in automated versus expert measurements of key features of their speech including MLU and question-asking. We have applied the automated pipeline to a dataset of classroom recordings from $13$ children and $3$ teachers observed on $17$ occasions over one year. This dataset contains $765$ hours of recordings and required $200$ hours of processing time with the current automated pipeline. The automated pipeline yielded over half a million transcribed utterances, which we are currently analyzing. Thus, automated methods show great potential for producing datasets with which to understand interaction and development in naturalistic contexts.

\section*{Acknowledgement}
This work was supported by the National Institute on Deafness and Communication Disorders (R01DC018542), the National Science Foundation (2150830), and the Simons Foundation Autism Research Initiative (SFI-AR-HUMAN-00004115-01).

\bibliographystyle{IEEEtran}
\bibliography{main}

\end{document}